\documentstyle[12pt,epsfig]{article}

\newcommand{\eq}{\begin{equation}}
\newcommand{\en}{\end{equation}}
\newcommand{\bea}{\begin{eqnarray}}
\newcommand{\eea}{\end{eqnarray}}
\newcommand{\spz}{\hspace{0.7cm}}

\newcommand{\ba}{\begin{array}}
\newcommand{\ea}{\end{array}}

\newcommand{\virg}{\spz,\spz}

\newcommand{\sts}{\footnotesize}
\newcommand{\scr}{\scriptsize}

\newcommand{\fr}{\rightarrow}

\newcommand{\ri}{\right}
\newcommand{\lf}{\left}
\newcommand{\th}{\theta}

\newcommand{\ep}{\varepsilon}

\newcommand{\D}{\Delta}

\newcommand{\ct}{\tilde{c}}
\newcommand{\Mt}{\tilde{M}}

\newcommand{\NP}[1]{Nucl.\ Phys.\ {\bf #1}}
\newcommand{\PL}[1]{Phys.\ Lett.\ {\bf #1}}

\newcommand{\MPL}[1]{Mod.\ Phys.\ Lett.\ {\bf #1}}
\newcommand{\IJMP}[1]{Int.\ J.\ Mod.\ Phys.\ {\bf #1}}

\begin{document}
\setlength{\unitlength}{1.5mm}
\newsavebox{\pole}
\sbox{\pole}
{\begin{picture}(2,2)(0,0)
\put(0,0){\line(1,3){1.5}}
\put(0,0){\line(-1,3){1.5}}
\put(0,4.5){\oval(3,3)[t]}
\end{picture}}
\newsavebox{\STNTBA}
\sbox{\STNTBA}{\begin{picture}(70,15)(0,-3)
\multiput(10,0)(9,0){2}{\circle{2}}
\put(11,0){\line(1,0){7}}
\put(19,1){\usebox{\pole}}
\put(10,-2){\makebox(0,0)[t]{{\protect\scr 0}}}
\put(19,-2){\makebox(0,0)[t]{{\protect\scr 1}}}
\end{picture}}
\newsavebox{\STNTBAA}
\sbox{\STNTBAA}{\begin{picture}(70,15)(0,-3)
\multiput(10,0)(9,0){3}{\circle{2}}
\multiput(11,0)(9,0){2}{\line(1,0){7}}
\put(19,1){\usebox{\pole}}
\put(10,-2){\makebox(0,0)[t]{{\protect\scr 0}}}
\put(19,-2){\makebox(0,0)[t]{{\protect\scr 1}}}
\put(28,-2){\makebox(0,0)[t]{{\protect\scr 2}}}
\end{picture}}
\newsavebox{\STNTBAB}
\sbox{\STNTBAB}{\begin{picture}(70,15)(0,-3)
\multiput(10,0)(9,0){3}{\circle{2}}
\multiput(11,0)(9,0){2}{\line(1,0){7}}
\multiput(19,1)(9,0){2}{\usebox{\pole}}
\put(10,-2){\makebox(0,0)[t]{{\protect\scr 0}}}
\put(19,-2){\makebox(0,0)[t]{{\protect\scr 1}}}
\put(28,-2){\makebox(0,0)[t]{{\protect\scr 2}}}
\end{picture}}

\renewcommand{\thefootnote}{\fnsymbol{footnote}}

\newpage
\setcounter{page}{0}
\begin{flushright}
Bologna preprint DFUB-93-12\\
Dec 1993 (enlarged Sept 1994)
\end{flushright}
\vskip 1cm
\begin{center}
{\bf INTEGRABLE PERTURBATIONS OF CFT WITH COMPLEX PARAMETER:\\
     THE $M_{3/5}$ MODEL AND ITS GENERALIZATIONS}\\
\vskip 1.4cm
{\large F.\ Ravanini$^1$, M.\ Stanishkov$^1$ and R.\ Tateo$^2$}
     \footnote{E-mail: ravanini@bologna.infn.it, stanishkov@bologna.infn.it,
     Roberto.Tateo@durham.ac.uk}\\
\vskip .5cm
{\em $^1$ I.N.F.N. - Sez. di Bologna, and Dip. di Fisica,\\
     Universit\`a di Bologna, Via Irnerio 46, I-40126 Bologna, Italy\\
\vskip .4cm
     $^2$ Dip. di Fisica Teorica, Universit\`a di Torino, Italy\\
     and Dept. of Mathematics, Durham Univ., Durham DH13LE, U.K.}\\
\end{center}
\vskip 1cm

\renewcommand{\thefootnote}{\arabic{footnote}}
\setcounter{footnote}{0}

\begin{abstract}
\noindent
We give evidence, by use of
the Thermodynamic Bethe Ansatz approach, of the
existence of both massive and massless behaviours for the $\phi_{2,1}$
perturbation of the $M_{3,5}$ non-unitary minimal model,
thus resolving apparent contradictions in the previous literature.
The two behaviours correspond to changing
the perturbing bare coupling constant from real values to
imaginary ones. Generalizations of this picture to the whole
class of non-unitary minimal models $M_{p,2p\pm 1}$,
perturbed by their least relevant
operator lead to a cascade of flows similar to that of unitary minimal models
perturbed by $\phi_{1,3}$. Various aspects and generalizations of this
phenomenon and the links with the Izergin-Korepin model are discussed.
\end{abstract}

\newpage

\section{Introduction}
The astonishing progresses of the last decade in two-dimensional conformal
Quantum Field Theory
(CFT), initiated with the work of Belavin, Polyakov and
Zamolodckikov~\cite{BPZ}, have developed in many directions. One of the most
interesting ones is the investigation of integrable perturbations of
CFT~\cite{Zam1}, where a lot of amusing non-perturbative phenomena can be
observed and the renormalization group flows can be reconstructed exactly in
many cases from the S-matrix of the corresponding Factorized Scattering Theory.

The bare action ${\cal A}$
of a CFT ${\cal M}$ perturbed by one of its relevant
operators $\Phi$ is given by
\eq
{\cal A} = {\cal A}_{\cal M} + \lambda \int d^2x \Phi(x)
\en
where ${\cal A}_{\cal M}$ is the (formal) action of the CFT. In what follows we
shall use the shorthand notation ${\cal M}+\Phi$ to refer to such a model.
The perturbing parameter $\lambda$, together with the parity of the perturbing
operator, plays a central role in determining the
possible behaviours (massive or massless) of the integrable theory.
For example, in the celebrated $\phi_{1,3}$
perturbations of unitary minimal models $(M_p + \phi_{1,3})$,
the perturbing operator is parity even
and the model is not necessarily equal for positive or negative $\lambda$.
Indeed, it shows a massive behaviour for negative $\lambda$ and a massless one
for positive $\lambda$, which have been widely studied in the literature.
There are other models, like the unitary and non-unitary minimal models $M_{p,
q}$ perturbed by their $\phi_{1,2}$ or
$\phi_{2,1}$ operators, as well as many other theories,
where the perturbing operator is odd, and this
implies an invariance for $\lambda \to -\lambda$ telling us that whatever the
behaviour for positive $\lambda$ is, it will be reproduced for negative
$\lambda$ too.

The interest of this paper is to clarify when a certain
integrable perturbation of a CFT admits both massive and massless behaviours.
The importance of discovering new
massless behaviours lies in the fact that they interpolate between
non-trivial ultraviolet (UV) and infrared (IR) CFT's,
thus providing very interesting information on the
structure of the two-dimensional Renormalization Group Space of Actions.
It has been widely believed
that the two behaviours can coexist only when there is no
symmetry $\lambda \to -\lambda$, i.e. when
the perturbing operator is even. When
this symmetry is present, instead, one could conclude at first glance that only
one of the two possible behaviours is allowed.
However, Fendley, Saleur and Al.Zamolodchikov~\cite{FSZ1,FSZ2}
have recently given
support to the possibility of a massless flow in the Sine-Gordon model where
the perturbing operator is $Z_2$-odd. This new type of flow is related
to an imaginary coupling constant in the potential. In this paper,
by examining the simple $M_{3,5}+\phi_{2,1}$ theory, we shall
learn that such a subtle situation can occur also in minimal models and other
rational CFT's, thus enlarging considerably the set of possible
integrable models. As a by product, this will also suggest the existence
of massless
flows with imaginary coupling constant (similar to those described
in~\cite{FSZ1,FSZ2}
for Sine-Gordon) in the Izergin-Korepin model.

The plan of this paper is as follows: in sect.2 we summarize some results about
the $M_{3,5}+\phi_{2,1}$ theory. Arguments in support of a massive behaviour
for this theory have been given in ref.~\cite{muss}, while a massless
interpretation has been supported in~\cite{mart}. The oddness of the $\phi_{2,
1}$ operator seems to imply a contradiction between the results of~\cite{muss}
and~\cite{mart}.

In sect.3, starting with the hypothesis that the model is
massive, with the S-matrix proposed in~\cite{muss}, we give a set of
Thermodynamic Bethe Ansatz (TBA) equations governing the Renormalization Group
(RG) evolution of the Casimir energy of the
vacuum and of the first excited state on a cylinder.  This is
obtained by folding the known set of TBA equations of a suitable $W_3$-minimal
model perturbed by its $\phi_{id, adj}$ operator.
The TBA set passes various checks that give a high level of confidence
to our conjecture.

Then, in sect.4 we turn our attention to the other
possibility, i.e. to the massless behaviour. We show that it is possible to
construct a self-consistent massless scattering theory interpolating the
ultraviolet (UV) $M_{3,5}$ model and the infrared (IR) $M_{2,5}$, the
celebrated
Lee-Yang singularity. As this scattering theory turns out to be diagonal, the
TBA equations can be deduced easily in this case. They correspond
exactly to the "massless version" of the previous massive TBA, according to the
empiric rule introduced by Zamolodchikov~\cite{Al4} to pass from massive to
massless TBA by substituting mass terms with left and right movers in a
suitably symmetric way in the TBA equations. The perturbative analysis of the
TBA solutions for the ground state Casimir energy leads to the solution of the
original puzzle: in this case it is the $\lambda\to i\lambda$ transformation
that leads from massive to massless regime.

Sect.5 deals with the generalization of the $M_{3,5}+\phi_{2,1}$ result to a
whole series of models $M_{p,2p\pm 1}$ perturbed by their least relevant
operator and the link of this series of models with the Izergin-Korepin
$A_2^{(2)}$ affine Toda Field Theory.
A straightforward generalization of the $M_{3,5}$ TBA equations leads
to the picture that for each of these models there exists both a massive
behaviour, for which the S-matrix should be deducible as a folding of a
suitable
$W_3$-minimal model one, and a massless regime of interpolating flows,
accompanied, like in the usual $M_p+\phi_{1,3}$ case, by a staircase that has
already been reported in~\cite{mart}. We also briefly discuss
why in this case, unlike the $M_p+\phi_{1,3}$ one,
large $p$ perturbation theory cannot be used to study the massless flow.

Our conclusions on these results are collected in the
final sect.6, where we collect
comments about other theories that could share the same phenomenon and
try to propose some criterion to select possible candidates.
Here we also briefly discuss some open issues and possible developments for
further research.

\section{The $M_{3,5}+\phi_{2,1}$ puzzle}
We begin our investigation by presenting an intriguing puzzle that appears
in the $\phi_{2,1}$ perturbation
of the $M_{3,5}$ non-unitary minimal model, one of the simplest CFT's
after the Lee-Yang singularity
($M_{2,5}$) and the critical Ising model ($M_{3,4}$).
The $M_{3,5}$ CFT has
central charge $c=-\frac{3}{5}$ and three non-trivial scalar primary fields
$\phi_{1,2}(x)$, $\phi_{1,3}(x)$ and $\phi_{2,1}(x)$
of (left) conformal dimensions $\Delta_{1,2}=-\frac{1}{20}$,
$\Delta_{1,3}=\frac{1}{5}$ and $\Delta_{2,1}=\frac{3}{4}$, which satisfy the
well known fusion rules of minimal models. The ground state of the theory does
not coincide with the conformal vacuum $|0\rangle$ as in unitary models,
but better with the state $|\Omega\rangle=\phi_{1,2}(0)|0\rangle$. The ground
state Casimir energy on a cylinder is therefore not proportional to the central
charge $c$, but to the so called {\em effective} central charge
$\ct=c-24\D_{1,2}=\frac{3}{5}$.

In the following we shall consider the model $M_{3,5}+\phi_{2,1}$, with action
\eq
{\cal A} = {\cal A}_{M_{3,5}} + \lambda \int d^2x \phi_{2,1}(x)
\label{action}
\en
that can be interpreted in Statistical Mechanics as the
thermal perturbation of the $Q=4\cos^2\frac{2\pi}{5}$ critical Potts model.
A counting argument \`a la Zamolodchikov and an explicit construction of
conserved currents for the
simplest cases shows that this model is integrable. Moreover, the operator
$\phi_{2,1}$ is odd, as one can see from the symmetries of the fusion rules and
therefore the model is invariant for $\lambda\to -\lambda$.

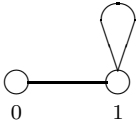
\begin{figure}[htbp]
\begin{center}
\begin{picture}(70,30)(0,0)
\put(20,5){\usebox{\STNTBA}}
\end{picture}
\vspace{-1.cm}
\caption{\label{fig1} \sts
The $T_2$ diagram describes the vacuum structure of the $M_{3,5}+\phi_{21}$
 model.}
\end{center}
\end{figure}

A factorizable S-matrix for the model (\ref{action}) can be proposed by quantum
reduction of the Izergin-Korepin model, as explained for general $M_{p,
q}+\phi_{2,1}$ in ref.~\cite{smirnov} and specialized to this simple case in
ref.~\cite{muss}. The model presents two vacua (call them 0 and 1). The
asymptotic states are identified with kink states $|K_{01}\rangle$ and
$|K_{10}\rangle$ interpolating them and a single state over the
1 vacuum $|K_{11}\rangle$. There is no similar state $|K_{00}\rangle$
over the 0 vacuum instead. The admissibility diagram of the corresponding
S-matrix is depicted in Fig.~\ref{fig1}, while the detailed expressions of the
S-matrix elements are given in ref.~\cite{muss}.
The author of ref.~\cite{muss}, after supporting his hypothesis
that the model is massive by a truncated conformal space analysis, concludes
that this is the only possible behaviour of the model, due to the symmetry
$\lambda\to -\lambda$.

However, $M_{3,5}+\phi_{2,1}$ belongs to a peculiar series of models,
that we shall denote below as $\tilde{M}_p+\psi$, ($p\geq 4$) where
\eq
\tilde{M}_p=\left\{ \begin{array}{ll}
M_{\frac{p+1}{2},p} & \mbox{for $p$ odd} \\
M_{\frac{p}{2},p+1} & \mbox{for $p$ even}
\end{array} \right.
\en
and
\eq
\psi=\left\{ \begin{array}{ll}
\phi_{2,1} & \mbox{for $p$ odd} \\
\phi_{1,5} & \mbox{for $p$ even}
\end{array} \right. = \mbox{least relevant operator of $\tilde{M}_p$}
\label{sei}
\en
The effective central charge of this series of models is given by
\eq
\ct=1-\frac{12}{p(p+1)}
\label{ctilde}
\en
{}From the Statistical Mechanics point of view, this series describes a
possible multicritical generalization of the Lee-Yang model,
definitely different from the one defined by a chain of usual $\phi_{1,3}$
perturbations and $\phi_{3,1}$ attractions.\footnote{The interpretation
of these multicritical versions of the Lee-Yang model as some {\em Lee-Yang
like singularities} related to each of the multicritical Ising points
corresponding to the unitary minimal series $M_p$ is not straightforward and
needs further investigation.}

In ref.~\cite{mart}, evidence is given for a staircase model related to the
$A_2^{(2)}$ Affine Toda Field Theory, which flows close to all models
$\Mt_p$. It is then natural, by analogy with all the other known examples of
staircase models~\cite{Al-staircase,Rav-Dorey1,Rav-Dorey2} to conjecture that
its $\theta_0\to \infty$ limit mimics a series of massless flows
\eq
\Mt_p + \psi \to \Mt_{p-1} ~\mbox{attracted by}~\left\{
\begin{array}{ll}
\phi_{2,1} & \mbox{for $p-1>4$ even} \\
\phi_{1,5} & \mbox{for $p-1$ odd} \\
T\bar{T} & \mbox{for $p-1=4$}
\end{array} \right.
\en
Notice that the attracting operators are the least irrelevant operators in the
$\Mt_{p-1}$ models. The situation looks very similar to the one known in
unitary minimal models where the perturbation by the least relevant operator
$\phi_{1,3}$ at
ultraviolet (UV) leads, for positive $\lambda$, to an IR point attracting the
flux by its least irrelevant operator $\phi_{3,1}$. The considerations of
ref.~\cite{lassig} should apply here similarly.

In particular, this supports the hypothesis of the existence of a massless
behaviour for the $M_{3,5}+\phi_{2,1}$ model, flowing towards an IR limit
governed by the $M_{2,5}$ model (the Lee-Yang singularity) with attraction
operator given by $T\bar{T}$. ($T$ and $\bar{T}$ denoting the two components of
the stress-energy tensor of $M_{2,5}$). The most dramatic point about this is
that the invariance $\lambda\to -\lambda$ should then imply that the model is
always massless, in striking contradiction with the results of
ref.~\cite{muss}.

In conclusion, the $M_{3,5}+\phi_{2,1}$ puzzle can be summarized as follows:
the $\lambda\to -\lambda$ invariance seems to predict only one possible
behaviour (massive or massless) for the $M_{3,5}+\phi_{2,1}$ model. However, in
the literature some reasonable evidence has been given to support both the
massive~\cite{muss} or massless~\cite{mart} interpretation. Who is right? Or,
is there a way out allowing both behaviours without contradiction? This is the
main question we shall clarify in the present paper.

\section{TBA for massive $M_{3,5}+\phi_{2,1}$}
As we said in the
previous section, a massive scattering theory has been proposed
for the model in question in ref.~\cite{muss}. The corresponding factorized
S-matrix is not diagonal and a direct Bethe Ansatz approach to deduce the
TBA equations governing the evolutions of energy
levels on a cylinder along the RG flow is very hard. One can however avoid this
difficulty by guessing the TBA equations and then verifying if they pass
various
checks to give them credibility. The starting point of our guesswork
is the observation that the central charge of the models
$\Mt_p$, eq.(\ref{ctilde}), is exactly half of that of the $(W_3)_p$
series of minimal models of $W_3$-invariant CFT~\cite{FatZam-w3}.
This suggests that a relation could occur between the
$(W_3)_p$ models perturbed by their $\phi_{id,adj}$ operator, $(W_3)_p +
\phi_{id,adj}$ for short, and our $\Mt_p + \psi$ integrable models.
The case of the bottom models of the two series, the
scaling Potts model ($(W_3)_4+\phi_{\Delta=2/5}$)
and the scaling Lee-Yang singularity
($M_{2,5}+\phi_{1,2}$) has been dealt with in ref.~\cite{Al1},
where it is
stressed that the folding procedure leading from the Potts S-matrix to the
Lee-Yang one induces an analogous folding in the structure of TBA equations.
The second models in the two
series are the $(W_3)_5 + \phi_{id,adj}$
and the $M_{3,5}+\phi_{2,1}$ models respectively.
The S-matrix of the former
is given in~\cite{DeVega-Fateev} and one
can verify against the explicit expressions of ref.~\cite{muss} that the same
kind of folding procedure works here too to produce the S-matrix of the latter.
It seems then natural to conceive that the TBA equations for the $M_{3,
5}+\phi_{2,1}$ model are those of the $(W_3)_5 + \phi_{id,adj}$ suitably
folded.

These latter are given in~\cite{Martins,Rav1} and read as follows (here $\star$
means convolution and $\Lambda_b^i=\log(1+e^{\varepsilon_b^i})$,
$L_a^j=\log(1+e^{-\varepsilon_a^j})$)
\eq
\nu_a^i = \varepsilon_a^i + \frac{1}{2\pi}
\phi \star \left[ \sum_{b=1}^2
G^{ab}(\nu_b^i-\Lambda_b^i) - \sum_{j=1}^2 H^{ij} L_a^j\right]
\virg i=1,2 \spz a=1,2
\label{A}
\en
where the kernel $\phi$ has the form
\eq
\phi(\theta) = \frac{3}{2\cosh\frac{3}{2}\theta}\spz ,
\en
the two matrices $G$ and $H$ have non-negative integer entries (and therefore
can be thought as incidence matrices of two graphs ${\cal G}$ and ${\cal H}$
respectively)
\eq
G=H=\left( \begin{array}{cc} 0 & 1\\ 1 & 0 \end{array}\right)
\spz\Longrightarrow\spz {\cal G}={\cal H}=A_2
\en
and the energy terms $\nu_a^i(\theta)$ are chosen as follows
\eq
\nu_a^1(\theta)=mR\cosh \theta \spz \nu_a^2 = 0 \virg a=1,2
\en
where $m$ is the mass of the fundamental kink of rapidity $\theta$
and $R$ the radius of the cylinder on which the theory is put.

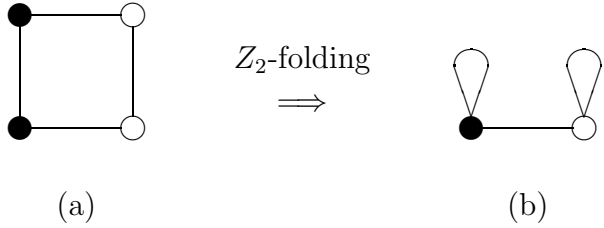
\begin{figure}
\begin{center}
\begin{picture}(70,25)
\multiput(10,10)(0,10){2}{\circle*{2}}
\multiput(20,10)(0,10){2}{\circle{2}}
\multiput(11,10)(0,10){2}{\line(1,0){8}}
\multiput(10,11)(10,0){2}{\line(0,1){8}}
\put(15,3){\makebox(0,0){(a)}}
\put(55,3){\makebox(0,0){(b)}}
\put(35,12){\makebox(0,0){$\Longrightarrow$}}
\put(35,16){\makebox(0,0){$Z_2$-folding}}
\put(50,10){\circle*{2}}
\put(60,10){\circle{2}}
\put(50,11){\usebox{\pole}}
\put(51,10){\line(1,0){8}}
\put(60,11){\usebox{\pole}}
\end{picture}
\caption{\label{fig2} \sts
Graphical representation of TBA systems encoded on ${\protect\cal G}
\protect\times {\protect\cal
H}$. The diagram ${\protect\cal G}$ develops vertically, ${\protect\cal H}$
horizontally: (a) the $A_2\times A_2$ case, corresponding to
$(W_3)_5+\phi_{id,adj}$; (b) the $T_1\times A_2$ case, corresponding to
$M_{3,5}+\phi_{2,1}$.
Nodes denoted by {\protect\large $\protect\circ$} are attached an energy
term $\nu_a^i\equiv 0$
(magnonic nodes), those denoted by {\protect\large $\protect\bullet$}
are attached $\nu_a^i=mR\cosh\theta$ (particle nodes).}
\end{center}
\end{figure}

This complicated set of equations can be more clearly depicted in graphical
form, as usually done by many authors (for details see e.g ref.~\cite{QRT}) by
drawing the tensor product graph ${\cal G}\times{\cal H}$ ($A_2\times A_2$
in our case), whose connectivity reproduces the coupling of variables in the
TBA eq.(\ref{A}) and giving to each node $(i,a)$ a choice for the energy term
$\nu_a^i$. Our case is depicted in Fig.~\ref{fig2}a.

The TBA eq.(\ref{A}) is symmetric in the $a$ indices. The aforementioned
folding procedure gives our conjecture for the TBA corresponding to the
S-matrix of ref.~\cite{muss}
\eq
\nu^i = \varepsilon^i + \frac{1}{2\pi}
\phi \star \left[ (\nu^i-\Lambda^i) - \sum_{j=1}^2 H^{ij} L^j\right]
\virg i=1,2
\label{B}
\en
where $\nu^1=mR\cosh\theta$ and $\nu^2=0$.
It can be interpreted in graphical form as the folding of the ${\cal G}=A_2$
diagram by its $Z_2$ symmetry to the $T_1$ diagram. Then this TBA set can be
depicted on the $T_1 \times A_2$ graph (see Fig.~\ref{fig2}b).

Now that we have our TBA equations at hand, we can evaluate various quantities
with them to learn more about the theory described by the S-matrix
of~\cite{muss}. The most immediate thing to compute is the scaling function
\eq
\ct(r)=\frac{3}{\pi^2}\int_{-\infty}^{+\infty}d\theta \nu^1(\theta) L^1(\theta)
\en
where $r=mR$ is a dimensionless scale, such that $t=\log r$ can be interpreted
as the RG {\em time} going from $-\infty$ at UV to $+\infty$ at IR.
In particular the UV limit $r\to 0$ can be exactly computed by resorting to
Dilogarithm identities, and results in $\ct_{UV}=\frac{3}{5}$ as expected.
Moreover, each solution of the TBA equations (\ref{B}) is also a
solution of the following system of functional equations
\eq
Y^i\left(\theta+\frac{i\pi}{3}\right) Y^i\left(\theta-\frac{i\pi}{3}\right)=
\frac{1+Y^i(\theta)}{\prod_{j=1}^2 (1+Y^j(\theta)^{-1})^{H^{ij}}}
\label{Y}
\en
The following periodicity property is shown by this system
\eq
Y^1(\theta+2\pi i)=Y^2(\theta)
\en
Al. Zamolodchikov~\cite{Al2} was able to relate this periodicity to the
conformal dimension of the UV perturbing operator. In our case this turns out
to give exactly $\Delta_{pert}=\frac{3}{4}$ as expected. These two fundamental
signals ($\ct_{UV}=\frac{3}{5}$ and $\D_{pert}=\frac{3}{4}$)
give us confidence in our conjecture, but many other interesting
properties can be analyzed. We try to summarize them in the following.

\subsection{Comparison with perturbation theory}

Around UV, the behaviour of the scaling function $\ct(r)$ is well known to be
summarized by the formula
\eq
\ct(r) = c_{UV} + \mbox{bulk} + \sum_{n=1}^{\infty} c_n (r^y)^n
\label{svil}
\en
where $y=2-2\D_{pert}=\frac{1}{2}$.
The non-perturbative bulk term that takes into account
long range fluctuations can be calculated from
general principles in the TBA approach~\cite{Al1,KM2,Al3,Al4,Al5}
and in the present case takes the form
\eq
\mbox{bulk} = \frac{3}{4\pi^2}r^2\log r
\en
The scaling function $\ct(r)$ can be computed numerically up to very high
precision.
Once the bulk term is subtracted, the $c_n$ coefficients can be estimated via
polynomial fit and
compared against the results of perturbation theory around the UV conformal
point. The UV perturbative series is
\eq
\ct_{pert} = c_{UV} + \sum_{n=1}^{\infty} P_n (\lambda R^{1/2})^n
\en
with the coefficients $P_n$ given by the UV correlation functions
\eq
P_n = 12\frac{(-1)^n}{n!}\sqrt{2\pi}
\int\prod_{j=1}^{n-1}\frac{d^2z_j}{(2\pi|z_j|)^{1/2}}\langle \Omega |
\phi_{2,1}(1,1)\prod_{j=1}^{n-1}\phi_{2,1}(z_j,\bar{z}_j)|\Omega\rangle
\en
Therefore the coefficients $c_n$ can be compared against the $P_n$ once the
proportionality constant $\kappa=\lambda/m^{1/2}$ is known
\eq
c_n = \kappa^n P_n
\label{prop}
\en
Recently Fateev~\cite{fat}
used external magnetic field techniques to calculate $\kappa$
independently of TBA for a large class of theories. His result in our case can
be summarized as
\eq
\kappa^2 = -{  m \over 3 \pi^2} { \Gamma\lf[ {5  \over 12} \ri]^2
\Gamma\lf[ {4  \over 3} \ri]^2 \over \Gamma\lf[ {7  \over 12} \ri]^2
\Gamma\lf[ {5  \over 3} \ri]^2 }
\en
thus providing the proportionality constant between $\lambda$ and $m^{1/2}$
\eq
\lambda=0.253001 i m^{1/2}
\label{ventuno}
\en

Surprisingly, for the first non-trivial coefficient $P_2$
(the $P_n$ with $n$ odd
are all zero by the symmetries of the
fusion rules) the check can be done even if $\kappa$ was not known. This is due
to the curious result that $P_2=0$ although the 4-point function needed to
calculate it is not zero:
\eq
P_2 = 6\int\frac{d^2z}{|z|^{1/2}} \langle \phi_{1,2}(\infty,
\infty) \phi_{2,1}(1,1) \phi_{2,1}(z,\bar{z}) \phi_{1,2}(0,0)\rangle
\en
This 4-point function involves only one conformal
block and its form can be fixed by using informations on its
monodromy and the $SL(2,\bf{C})$ invariance. As a final result one is led to
compute a Dotsenko-Fateev like integral whose quadrature is identically zero
thanks to the property of the hypergeometric function $F(-1,b;-b;-1)=0$,
$\forall b$.

What we have done was first to fit $(\ct(r)-\mbox{bulk})$
with a polynomial having both
even and odd powers of $r^{1/2}$, to check that really the $c_n$ with $n$
odd are all zero within numerical error. Then
we turned to a fit with only even powers of $r^{1/2}$ to make more precise
estimates
of the $c_{2n}$. The results are collected in the first column of
table~\ref{tab1}, where it appears clearly that the $c_2$ coefficient is zero
within an approximation of $10^{-14}$. Table~\ref{tab1} contains also other
information that will become clear in the next pages.

\begin{table}
\begin{center}
\begin{tabular}{||c|c|c|c||}                          \hline \hline
$c_n$ &   massive               &   massless               &$ \Delta c$  \\
\hline
$c_0$ & $ 0.6 $                  & $0.6$                   & /           \\
\hline
$c_2$ & $-5.4757325311081032\cdot 10^{-14}$ &
$-8.8531745297863157\cdot 10^{-15}$&$\pm 6\cdot 10^{-14}$ \\
\hline
$c_4$ & $-0.1755364844390       $&$-0.1755364844395$       &
$\pm 7\cdot 10^{-13}$ \\
\hline
$c_6$ & $ 2.2294559673\cdot 10^{-2}      $&
$-2.22945596576\cdot 10^{-2}$     &$\pm 4\cdot 10^{-12}$ \\
\hline
$c_8$ & $-1.012291\cdot 10^{-3}          $&
$-1.012290\cdot 10^{-3}$          &$\pm 2\cdot 10^{-9}$  \\
\hline
$c_{10}$ & $-8.6210\cdot 10^{-4}            $&
$ 8.6212\cdot 10^{-4}$            &$\pm 2\cdot 10^{-8}$  \\
\hline
$c_{12}$ & $-9.87\cdot 10^{-5}              $&$-9.85\cdot 10^{-5}$
&$\pm 3\cdot 10^{-7}$  \\
\hline \hline
\end{tabular}
\parbox{130mm}{\caption{\label{tab1}}
\protect{\footnotesize Massive and massless UV coefficients. The last column
reports the difference between the absolute values of the two.} }
\end{center}
\end{table}

\subsection{TBA  for the first excited state}

\begin{figure}
\begin{center}
\vspace{-1.cm}
{}~\epsfig{file=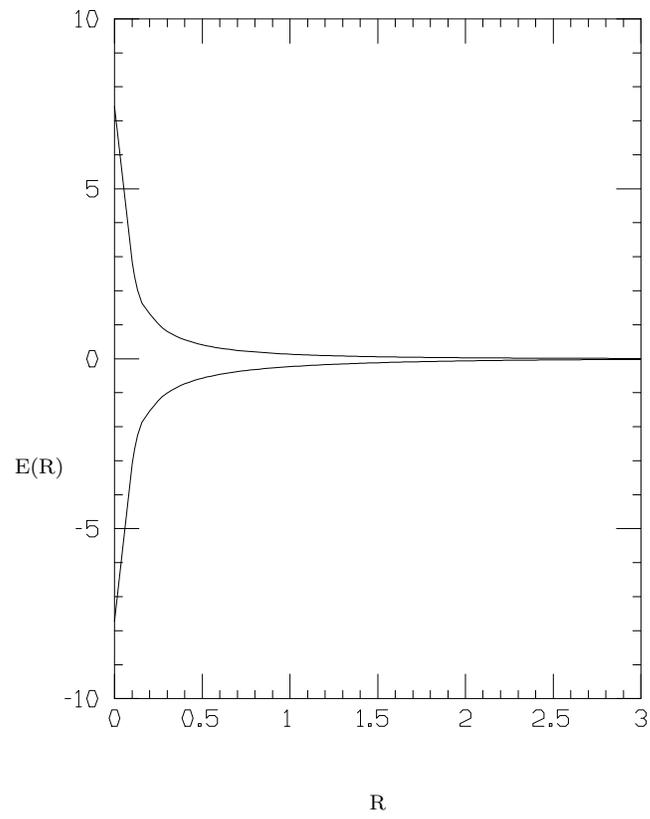,width=10cm}
\put(-60,40){\makebox(0,0)[t]{{\protect\scr E(R)}}}
\put(-30,10){\makebox(0,0)[t]{{\protect\scr R}}}
\vspace{-1.5cm}
\caption{\label{fig3}
\sts The vacuum  and the first excited states obtained using the TBA
system.}
\end{center}
\end{figure}

The massive TBA system in presence of a chemical
potential (with suitable chioces of the latter)
is known to yield the behaviour of various excited
states~\cite{KM3,Fendley}.
We find that the TBA system obtained from (\ref{B})
with the substitution
\eq
e^{-\ep(\th)} \fr -e^{-\ep(\th)}
\label{sub}
\en
in $\Lambda^i(\theta)$ and $L^i(\theta)$,
describes the first excited state of the perturbed massive theory. The UV limit
of this state is the conformal vacuum $|0\rangle$, therefore its Casimir energy
is directly proportional to the central charge of the
$M_{3,5}$ model. The
substitution~(\ref{sub}) does not influence the periodicity property of
$e^{-\ep(\th)}$ and in the UV regime we have a perturbative expansion
like eq.~(\ref{svil}). At $r\to 0$ we obtain the central charge of the
theory via standard dilogarithmic sum-rules. We find $c_{UV}=-\frac{3}{5}$
as it must be for the model $M_{3,5}$.
We also solve
numerically this TBA system and in Fig.~\ref{fig3} we report, as functions of
$R$, the first
two energy levels obtained from the numerical solution of eqs.
(\ref{B}) for the ground state energy $E_0(R)$ and with the substitution
(\ref{sub}) for the first excited state energy $E_1(R)$.
We see that the two
levels exponentially degenerate in agreement with the double well
potential of the theory as interpreted in~\cite{muss}.

\begin{figure}
\begin{center}
\vspace{-1.cm}
{}~\epsfig{file=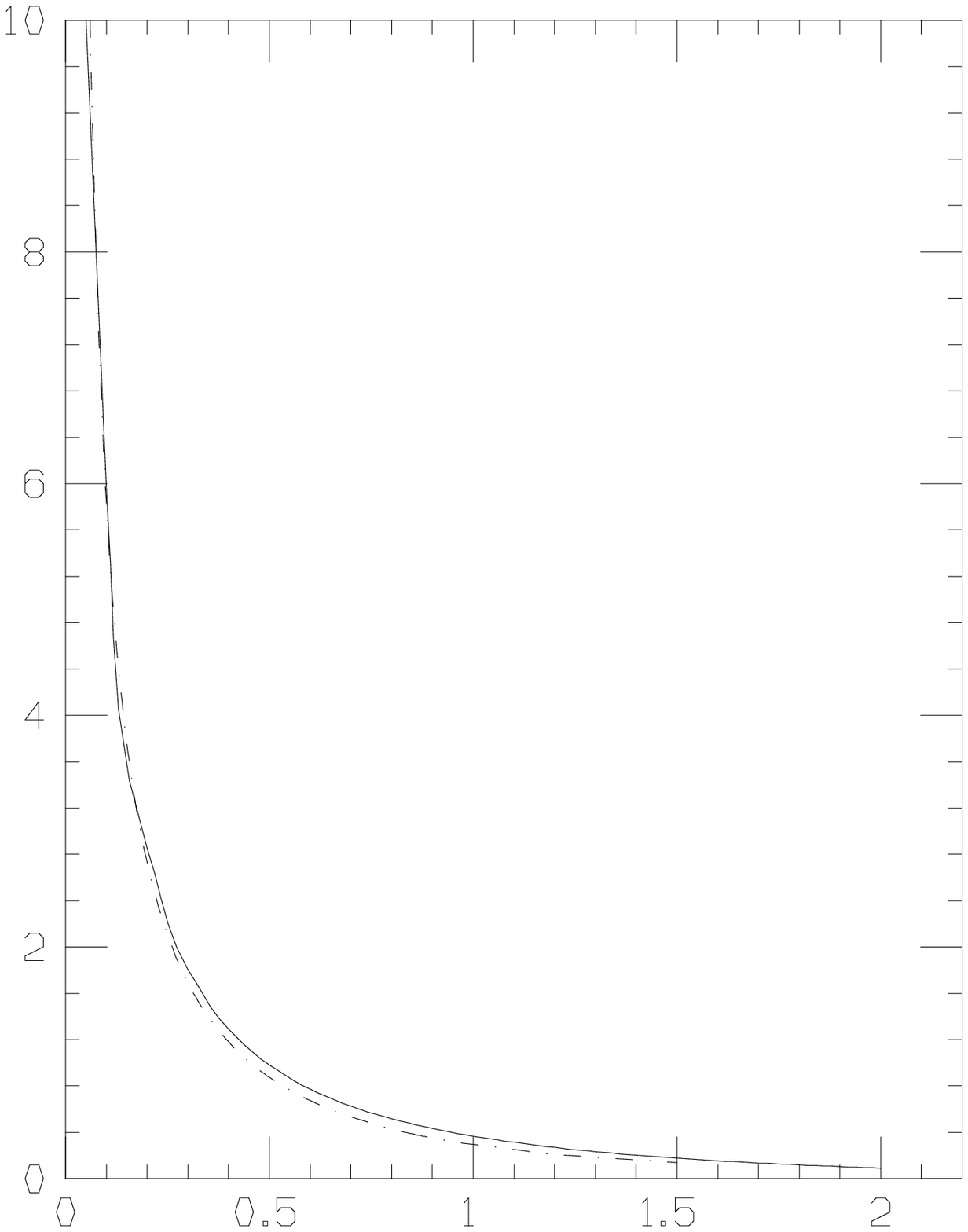,width=10cm}
\put(-60,40){\makebox(0,0)[t]{{\protect\scr$\Delta E(R)$}}}
\put(-30,10){\makebox(0,0)[t]{{\protect\scr R}}}
\vspace{-1.5cm}
\caption{\label{fig4}
 \sts The energy gap between the vacuum and the first excited
state decaying exponentially with the volume R. We compare the gap shape
obtained using the TBA with that obtained using TCSA (dotted line).}
\end{center}
\end{figure}

In Fig.~\ref{fig4} we represent the function
$\Delta E(R)= E_1(R)-E_0(R)$ compared with the result obtained in ref.
{}~\cite{muss} using the Truncated Conformal Space Approach (TCSA).
Up to a little discrepancy
due to the finite level truncation of the Hilbert space in TCSA, we
see that the two results agree, thus giving another piece of strong support to
the TBA system (\ref{B}).

In order to give even more support to the massive TBA we check the first
numerical coefficient against perturbation theory along the same lines of the
previous sub-section. In this case the value for $\kappa$ is needed and gives
\eq
c_1= \kappa P_1 = { 8 \over \pi }
{
\Gamma\lf[ {5  \over 12} \ri]^2
\Gamma\lf[ {3  \over 4} \ri]^2
\Gamma\lf[ {4  \over 3} \ri]^2
\over
\Gamma\lf[ {1  \over 4} \ri]^2
\Gamma\lf[ {7  \over 12} \ri]^2
\Gamma\lf[ {5  \over 3} \ri]^2
} \sim 0.551329
\en
Using standard numerical solution of
the excited TBA system we find $c_1= 0.5513 \pm 0.0001$, in very good
agreement with the perturbative result.


\subsection{IR regime}

For massive TBA systems, we may compare the IR behavior predicted by TBA
equations and the one predicted by general considerations (cluster expansions)
on the associated scattering theory. For kink like scattering,
these latter predict that the leading contribution to
$E(R)$ is proportional to the integral
\eq
I(r)= - {1 \over 2 \pi} \int \cosh \theta e^{-r \cosh \theta} d \theta
\en
multiplied by the largest eigenvalue $\Lambda_{max}$
of the kink adjacency matrix, as discussed in ref.~\cite{Al3}.
So  the leading term of the IR asymptotic is
\eq
E_0(R) \sim \Lambda_{max} m I(r)
\en
The vacuum structure of our theory is described by the $T_2$ diagram of
Fig.~\ref{fig1}. Its incidence matrix has eigenvalues
\eq
\Lambda_{max}= 2 \cos \lf( {  \pi \over 5} \ri) \virg
\Lambda_{min}= 2 \cos \lf( {3 \pi \over 5} \ri)
\en
Our TBA system (\ref{B}) in this limit predicts
\eq
\ep(\th) \sim
r \cosh \th - \log \lf( 1+e^{-\ep(\infty)} \ri)=
r \cosh \th - \log \lf[ 2 \cos\lf( {\pi \over 5} \ri) \ri]
\en
and
\eq
E_0(R) \sim - {m \over 2 \pi} \int d \theta \cosh \theta e^{-\ep(\theta)}=
 2 \cos \lf( {  \pi \over 5} \ri) m I(r)
\en
in agreement with the expected result.
The lowest excited state should become degenerate with the ground state in
infinite volume, from
our TBA system modified by (\ref{sub}) for the excited state we find
\eq
\ep(\th) \sim
r \cosh \th - \log \lf( e^{-\ep(\infty)}-1 \ri)=
r \cosh \th - \log \lf[ -2 \cos\lf( {3 \pi \over 5} \ri) \ri]
\en
and
\eq
E_1(R) \sim - {m \over 2 \pi} \int d \theta \cosh \theta e^{-\ep(\theta)}=
2 \cos \lf( { 3 \pi \over 5} \ri) m I(r)
\label{ecci}
\en
This result confirms that  the splitting $\D E(R)$ decays exponentially
with the volume $R$. Notice that the coefficient in (\ref{ecci})
corresponds to
the second eigenvalue of the incidence matrix of $T_2$. This
seems to be a quite general property of the degenerate vacuum
state theories and  suggests the existence of some kind of general
relation between TBA and its graph encoding
\cite{QRT} and the usual graph theory of S-matrix adjacency. This
investigation is in progress.

\section{Massless Scattering for $M_{3,5}+\phi_{2,1}\to M_{2,5}$}

In this section, we wish to show that it is possible to propose a massless
scattering theory describing a flow from the UV
$M_{3,5}+\phi_{2,1}$ down to the IR $M_{2,5}$ model.
The S-matrix formalism can be developed for a massless scattering theory
following the lines proposed by Alexander and Alexey Zamolodchikov
in~\cite{ZamZam2} and developed in~\cite{FSZ2,FS}. We refer the reader to those
papers for a treatment of the subtelties concerning the definition of a
massless S-matrix and the properties related to it. In particular the deduction
of massless S-matrix for the minimal models described in~\cite{FSZ2} will be
followed quite closely. A massless scattering is
defined as consistency requirement of the Bethe equations, thus overcoming
kinematic difficulties. Instead of massive asymptotic
states (particles), one introduces right ($+$) and left ($-$) movers whose
energy can be parametrized as $\frac{m}{2}e^{\pm\theta}$ respectively
($m$ is a scale of the theory). Four scattering matrices have to be
considered: $S_{++}$, $S_{--}$, $S_{+-}$ and $S_{-+}$. Parity invariance
requires that $S_{++}=S_{--}$ and by analogy to the $M_p+\phi_{1,3}$ case we
also require $S_{-+}=S_{+-}$. $S_{++}$ is scale invariant
and can be thought as the scattering matrix for the left sector of the IR CFT
($M_{2,5}$ in our case). Analogy with the unitary minimal model series suggests
to take it formally equal to the S-matrix of the IR model perturbed by its
least
relevant operator in the massive direction. Therefore, let us consider the
massless scattering defined by taking
$S_{++}=S_{LY}$, where
$S_{LY}$ denotes the S-matrix of the massive
perturbation of the
$M_{2,5}$ model by its operator of conformal dimension $-\frac{1}{5}$ given
in~\cite{Cardy-Muss}. i.e. let us choose
\eq
S_{++}=S_{--}=-\left(\frac{1}{3}\right)\left(\frac{2}{3}\right)
\label{LL}
\en
We use here the notation $(x)=\frac{\sinh\frac{1}{2}(\theta-i\pi x)}
{\sinh\frac{1}{2}(\theta-i\pi x)}$. By analogy with what assumed in
ref.~\cite{FSZ2}, we take $S_{+-}(\theta)$
proportional to $S_{++}(\theta-i\alpha)$, for some rapidity shift
$\alpha$, and $S_{-+}\propto S_{++}(\theta+i\alpha)$. A consistent choice,
compatible with the requirement $S_{-+}=S_{+-}$, is to fix $\alpha=\pi/2$ and
the proportionality factor equal to 1. Therefore
\eq
S_{+-}(\theta)=-\left(-\frac{1}{3}\right)\left(-\frac{2}{3}\right)=
\left(S_{LY}\right)^{-1}
\label{LR}
\en
We propose as scattering theory corresponding to the massless $M_{3,5}+\phi_{2,
1}$ model the one governed by the $S_{++}$ and $S_{+-}$ described above.

\subsection{TBA and UV behaviour}
Of course this proposal must be checked
by computing the UV and IR behaviour of the
theory defined by this S-matrix and showing that UV really gives $M_{3,5}$
perturbed by $\phi_{2,1}$ and IR gives $M_{2,5}$ with attraction operator
$T\bar{T}$. This can be done by resorting again to the TBA approach.
The deduction of the TBA system from the S-matrix is
particularly simple here, as the latter is diagonal.
Standard calculations give the TBA equations for the scattering theory defined
by (\ref{LL},\ref{LR}) in the form
\eq
\frac{mR}{2}e^{\pm\theta}=\varepsilon_{\pm}(\theta)+\frac{1}{2\pi}
[\hat{\phi}\star (L_{\pm}-L_{\mp})](\theta)
\label{TBA}
\en
where $R$ is the radius of the cylinder and $L_{\pm}$
is short for $\log(1+e^{-\varepsilon_{\pm}})$.
The convolution kernel $\hat{\phi}(\theta)$ is given by
\eq
\hat{\phi}(\theta)=
-i\frac{d}{d\theta}\log S_{++}(\theta)=i\frac{d}{d\theta}\log S_{+-}(\theta)
\en
Simple manipulations involving a Fourier transform, bring this TBA system into
one having the same form of eq.(\ref{B}), where now $i=+,-$.
This massless TBA is therefore
encodable, like the massive one of the previous section, on the $T_1\times A_2$
graph. The only difference is in the choice of the energy terms attached to the
two nodes. Here they read $\nu^{\pm}=\frac{r}{2}\exp (\pm\theta)$.
The scaling function
\eq
\ct (r)=\frac{3}{\pi^2}\frac{r}{2}\int d\theta [e^{\theta}L_+(\theta) +
e^{-\theta}L_-(\theta)]
\en
can easily be evaluated in the two $r \to 0$ (UV) and $r\to\infty$ (IR) limits
by resorting to Rogers Dilogarithm sum rules and gives
$\ct_{UV}=\frac{3}{5}$ and $\ct_{IR}=\frac{2}{5}$. The TBA system
(\ref{TBA}) can also
be recast in the form of a set of functional equations equal to that of eq.
(\ref{Y}) (we just pick up a different solution with another asymptotic
condition on the pseudoenergies $\varepsilon_a$),
thus predicting from the already mentioned periodicity that the UV
perturbing operator has conformal dimension $\D_{pert}=\frac{3}{4}$.
The stationary (i.e. independent of $\theta$) version of the Y-system is a set
of algebraic equations whose solutions, once inserted in Dilogarithm sum rules,
not only predict the UV effective central charge, but also the
conformal dimensions of other excited states
and even subsets of the CFT fusion rules~\cite{Nahm}. Such
an exercise applied to the present case confirms with no doubt that the UV
limit of the scattering theory (\ref{LL},\ref{LR}) is the $M_{3,5}$ minimal CFT
perturbed by its $\phi_{2,1}$ relevant operator.

\subsection{IR behaviour}
What is different from the previous
massive case is the $r\to\infty$ behaviour near IR. This latter is dictated by
the Y-system with the right mover deleted, i.e. by the $T_1\times A_1$ system
encoding the effective central charge, conformal dimensions and fusion
rules~\cite{Nahm} of the $\ct=\frac{2}{5}$ Lee-Yang singularity model $M_{2,
5}$. The Y-system periodicity at IR must be compared against the asymptotic
expansion in $r^{-2(1+\D_{attr})}$~\cite{Al4},
where $\D_{attr}$ is the
conformal dimension of the irrelevant operator of the $M_{2,5}$ CFT attracting
the flow. It turns out that $\D_{attr}=2$, thus confirming the identification
of such operator with $T\bar{T}$, as expected.

Results of perturbation theory around the IR Conformal
point for the massless TBA (see ref.~\cite{KM3})
are described by the asymptotic perturbative series
\eq
\ct_{pert} \sim -\sum_{n=0}^{\infty} b_n \lf( {- \pi^3 g \over 6 R^2} \ri)^n
\label{ex}
\en
The  first $b_n$ coefficients are
\eq
b_0= (12 \Delta_{min} - c_{IR}) \virg b_1= -(12 \Delta_{min} -c_{IR})^2 \virg
b_2= 2(12 \Delta_{min}- c_{IR})^3
\en
We must compare the perturbative expansion~(\ref{ex}) with
the numerical solution of our TBA system
\eq
\ct_{pert} \sim -\sum_{n=0}^{\infty} \tilde{b}_n (R M)^{-n}
\label{ex2}
\en
The first three coefficients are
\eq
\tilde{b}_0= { 2 \over 5}
\en
(obtained using the dilogarithm sum-rule) and
\eq
\tilde{b}_1=0.58041579  \pm 8\cdot 10^{-8}\virg
\tilde{b}_2= 1.68440 \pm 2\cdot 10^{-5}
\label{coe}
\en
Setting $ K g= m^{-1}$ we use the second coefficient in order to fix the
constant $K$ ,we find $K={25 \over 4}\tilde{b}_1 = 3.627598 \pm 2\cdot 10^{-6}$
thus allowing to check the third
coefficient. We find $\tilde{b}_2=
 2 K^2 \lf ( 2 \over 5 \ri)^3 =1.68441 \pm 2\cdot 10^{-5}$
in good numerical agreement with (\ref{coe}).
All these checks confirm that the IR limit of the scattering theory
(\ref{LL},\ref{LR}) is the $M_{2,5}$ Lee-Yang CFT,
with $T\bar{T}$ attraction operator.

\subsection{Analytic continuation and solution of the puzzle}
Having now a quite solid ground of evidence for both massive and massless
regimes, we turn our attention to the main question of the paper: how can the
two behaviours coexist, in spite of the $\lambda\to -\lambda$ invariance? To
give an answer, first of all
notice that the TBA systems (\ref{B}) and (\ref{TBA}) are
a perfect example of the recipe observed in a
lot of integrable perturbed CFT having both massive and massless regimes, where
one passes from massless to massive TBA by keeping the
same structure of the equations, and modifying
the energy terms only. Instead of the right mover $\frac{r}{2}e^{\theta}$ one
puts a massive particle energy term $r\cosh \theta$, and instead of the left
mover $\frac{r}{2}e^{-\theta}$ one replaces a magnonic term with energy 0. This
procedure is well known in unitary minimal models perturbed by $\phi_{1,
3}$~\cite{Al3,Al4}, as well as for many coset generalizations of
them~\cite{Al5,Rav1} and also on other examples~\cite{FatAl,RTV,QRT}.
This observation is also expected to be equivalent to relate the massive and
massless scaling functions by analytic
continuation in the parameter $\lambda$~\cite{Al3}.
For the Triciritical Ising model examined in~\cite{Al3}, as well as for all
$M_p+\phi_{1,3}$, the analytic
continuation was from positive to negative $\lambda$. Here we know this cannot
be the case, due to the $\lambda\to -\lambda$ invariance. To have an idea of
what kind of analytic continuation is needed, we just examine the perturbation
theory around UV. For the massive case this has already been done in section
3.1, for the massless case the calculations repeat exactly the same structure.
The non-perturbative
bulk term, due to the different asymptotics chosen
on the TBA diagram, is not the same as in the massive case, but rather reads
\eq
\mbox{bulk} = \frac{\sqrt{3} r^2}{4\pi} + \frac{3}{4\pi^2}r^2 \log r
\en
Notice that in addition to the logarithmic contribution of the
massive case we have here also a contribution proportional to $r^2$, more
similar to those discussed in~\cite{Al1,KM2}.

Once this bulk term is subtracted, the numerical criteria as in the massive
case can be adopted here to extract the coefficients $c_n$
(see eq.(\ref{svil})) listed in the second column of table~\ref{tab1}.
Observe that, within numerical error, one passes from massive $\ct(r)$ to
massless one just by readjusting the coefficients as
\eq
c_n \to (-1)^{n/2} c_n
\en
The $P_n$ coefficients computed from CFT correlators
must be strictly the same, which means, in view of the relation (\ref{prop})
between $c_n$ and $P_n$, having the following recipe in passing from massive to
massless regime
\eq
\lambda \to i\lambda
\label{L}
\en
The scaling function (and therefore the ground state Casimir energy) remain
real after the substitution (\ref{L}), as it is (at least perturbatively) an
even function of $\lambda$, i.e. depends only on $\lambda^2$.
Thus it is natural to propose (\ref{L}) as the
analytic continuation allowing to pass from massive to massless regime. Notice
that in view of eq.(\ref{ventuno}), it is the massless theory that has real
$\lambda$, while the massive flow develops on the imaginary $\lambda$ axis.

The analytic continuation now proposed can be checked even beyond the
convergence radius of perturbation theory by resorting to the numerical
technique of Pad\'e approximants. Although we have only done some rough checks
with this method, they agree, within numerical error, with the data
for $\ct(r)$ extracted from the two TBA's.

The solution of the puzzle of sect.2 is then that both papers~\cite{muss}
and~\cite{mart} report part of the truth, in the sense that both behaviours are
allowed from a physical point of view. For both of them it is possible to
propose a scattering theory and a TBA system driving the RG behaviour from UV
to IR. What is new in the point of view of this paper is to allow the
perturbing parameter $\lambda$ to take in principle any complex value (which
does not affect the integrability properties of the model) and then check if
there is some direction in the complex $\lambda$ plane where one can still
define a consistent theory having real energies for all the states. In the case
$M_{3,5}+\phi_{2,1}$ examined here
we observe that the $Z_2$ symmetry implemented by $\lambda\to -\lambda$
implies that all observables are even or odd functions of $\lambda$. Moreover,
in a non-unitary theory like $M_{3,5}+\phi_{2,1}$ it is not necessary that the
action (\ref{action}) must be real (which, as the operator $\phi_{2,1}$ is
self-conjugate, would have implied reality of $\lambda$). For example, the
transformation $\lambda\to i\lambda$ keeps all even observables unchanged and
trivially multiplies all odd observables by an $i$ prefactor that can be
readsorbed in the normalization. The theory defined by $i\lambda$ is then still
a consistent one.

We think that the behaviour observed in this simple theory is just an example
of a phenomenon with a much wider scope, that can enlarge the zoo of
integrable models by a lot of interesting new cases. This is somewhat parallel
to the direction of investigation pointed out recently by
ref.~\cite{FSZ1,FSZ2}.
In next sections we start exploring this larger zoo by discussing some other
models sharing this phenomenon.

\section{The series $\tilde{M}_p+\psi$ and the Izergin-Korepin model}

In this section we generalize the results of Sect.3 and 4 to
the whole class of models $\Mt+\psi$.
First of all, let us observe that all these
models can be seen as quantum group reductions of the Izergin-Korepin model
(the $A_2^{(2)}$ Affine Toda Field Theory)~\cite{smirnov}.
The latter has Lagrangian
\eq
L= {1 \over 2} (\partial_{\mu} \phi)^2 + {1 \over \beta^2} e^{i \beta \phi}
- {\lambda \over \beta^2} e^{-i { \beta \over 2} \phi}
\label{liu}
\en
and can be seen as a complex Liouville theory perturbed by the field
$V(\phi)=e^{-i { \beta \over 2} \phi}$.
After quantum group reduction, we can consider the system (\ref{liu}) at
$\frac{\beta^2}{8\pi}={ p \over q}$
as a perturbed minimal theory $M_{p,q}$. In this picture the field
$e^{- i { \beta \over 2} \phi}$  is the
perturbing $\phi_{1,2}$ operator. At quantum level, by
interchanging the role of $p$ and $q$ the field $V(\phi)$
can also be associated to the $\phi_{2,1}$ operator. The
theory (\ref{liu}) can describe both these two different
integrable perturbations. This seems to provide only half of the models
$\Mt_p+\psi$, those where $\psi=\phi_{2,1}$. However also the other ones with
$\psi=\phi_{1,5}$ can be described in the same framework. One just has to
change the role of the two vertex operators in (\ref{liu}) and send
$\beta\to 2\beta$ to have
\eq
L= {1 \over 2} (\partial_{\mu} \phi)^2 - {\lambda \over \beta^2}
e^{-i 2\beta \phi}
+ {1 \over \beta^2} e^{+i \beta \phi}
\label{liu1}
\en
With this simple modification  at $\frac{\beta^2}{8\pi}={ p \over q}$
we can associate the Lagrangian (\ref{liu1}) to the perturbation of the
minimal theory $M_{p,q}$ by the operator $\phi_{15}$. This operator, as the
$\phi_{21}$ operator, is relevant
only in particular minimal theories, systematic check using the counting
argument reveals that  the field $\phi_{15}$ is always integrable
but it is  relevant only for $2p < q$, condition which is always
satisfied by the models $\Mt_p$ with $p$ even.

We prefer to keep the interesting problem to write the (massive and massless)
S-matrices for this series of models out of the scope of this paper.
Here we just observe that they could be
deduced both as reductions of the Izergin-Korepin S-matrix (the problems with
real analyticity in~\cite{smirnov} seem to be circumvented in
ref.~\cite{koubek}), or
as foldings of the $W_3$-minimal model ones. The matching between the two
formulations of the problem should give an interesting check on the ideas of
this section.

A simple proposal
for a TBA system for these models can be made by just adding more and more
magnonic nodes to the TBA of the $M_{3,5}+\phi_{2,1}$ model. This is compatible
with the folding of $W_3$ minimal models mentioned in sect.3, and with the
procedure usually adopted in minimal model series~\cite{Al3,Al4} as well as in
many series of rational CFT perturbed by $\phi_{id,adj}$ operators~\cite{Rav1}.
The TBA system for the model $\Mt_p+\psi$ is written exactly as eq.(\ref{B}),
with the sum now running from 1 to $k=p-3$ and can be encoded on a $T_1\times
A_k$ Dynkin diagram. The massive regime will be described by the chioce of
energy terms with the first node as $\nu^1=r\cosh\theta$ and all the others
$\nu^i=0$. The massless behaviour, instead, is reproduced by choosing
$\nu^1=\frac{r}{2}e^{\theta}$, $\nu^k=\frac{r}{2}e^{-\theta}$ and all others
$\nu^i=0$. This is very similar to the TBA structure for unitary minimal models
perturbed by $\phi_{1,3}$ introduced in~\cite{Al3,Al4}. Of course Dilogarithm
sum rules predict the expected UV and IR effective central charges and the
Y-system periodicities are compatible with the conformal dimension of the
perturbing operators as well as the attracting ones. In particular the massless
regime consists in a series of flows hopping from $\Mt_p$ to $\Mt_{p-1}$,
exactly as described in Sect.2, eq.(\ref{sei}). As in all known cases of series
of hopping flows, it is not surprising that this is accompanied by a staircase
model, which is the one described in ref.~\cite{mart}.

To give increased evidence to this picture, we have checked numerically the two
cases next to the $M_{3,5}+\phi_{2,1}$ model, namely $M_{3,7}+\phi_{1,5}$ and
$M_{4,7}+\phi_{2,1}$. In the first case the perturbing operator is even, the
scaling function has an expansion both in even and odd powers of $\lambda$ and
it is not difficult to check that passing from massive to massless behaviours
amounts to sending $\lambda\to -\lambda$.
This is common to all $p$ even models.
In the $p$ odd cases we are in the same situation examined earlier in the
paper: it is the $\lambda\to i\lambda$ transformation that leads from massive
(imaginary $\lambda$) to massless (real $\lambda$) behaviour.

\subsection{Non-perturbative nature of the flows}
One could ask why we do not check the existence of these massless flows in a
perturbative framework at large $p$, like it has been done in unitary minimal
models perurbed by $\phi_{1,3}$~\cite{Zam-pert,Ludwig-Cardy}. The reason is
that perturbative calculation always fails to pick up the IR fixed point in our
case, as it turns out from the following analysis.

The perturbative expansion of the two-point function $\langle \psi(x) \psi(0)
\rangle$ around the conformal point is given by
\bea
\langle \psi(x)\psi(0) \rangle &=&
\langle \psi(x)\psi(0) \rangle_{CFT} + \lambda
\int d^2y \langle\psi(x)\psi(0)\psi(y)\rangle_{CFT} \\
&+& \frac{1}{2}\lambda^2
\int d^2yd^2w\langle\psi(x)\psi(0)\psi(y)\psi(w)\rangle_{CFT} + ...
\label{expan}
\eea
Inserting this in the Callan-Symanzik equation provides a perturbative
definition of the anomalous dimension of the field $\psi(x)$ and hence the
perturbative $\beta$-function~\cite{Zam-pert,Capp-LaT}.

Consider first the models with $p$ even: $\psi=\phi_{1,5}$. For
$p\to\infty$ the parameter controlling the perturbation theory is
$\varepsilon=1-\D_{1,5}=\frac{3}{p+1}$. The first nontrivial
contribution comes from the order $\lambda$ in the expansion (\ref{expan}). The
integration is done explicitly in this case leading to the following
expression for the $\beta$-function
\eq
\beta(g)=\varepsilon g - \pi C g^2 + O(g^2)
\label{beta}
\en
$C$ is the structure constant of the channel $\phi_{1,5}\phi_{1,5}\to C\phi_{1,
5}$ and $g$ the renormalized coupling constant.
Surprisingly, unlike the case of unitary minimal models perturbed by
$\phi_{1,3}$, one finds here that $C=\frac{9}{4}\varepsilon + O(\varepsilon^2)
$. We are interested in finding a non-trivial fixed point $g^*$ in the
perturbative region around $g=0$, i.e. $g^* \sim \varepsilon$.
However, formal inserting of the
value for $C$ in (\ref{beta}) leads to a "nontrivial fixed point"
\eq
g^* = \frac{4}{9}\pi \sim O(1)
\en
which is obviously out of the range of perturbation theory.

Let us now turn to the second case $p$ odd ($\psi=\phi_{2,1}$). As it was
mentioned above, the operator $\phi_{2,1}$ is $Z_2$-odd and hence its 3-point
function vanishes identically. Thus, the first nontrivial contibution here
comes from the second order, i.e. the 4-point function. We analyse the leading
term of the channel $\phi_{2,1}\phi_{2,1}\to D\phi_{3,1}$. This gives
\eq
I_2 = \frac{1}{2}\lambda^2 D^2 (x^2)^{-2\D+\varepsilon}\int d^2w
|w|^{2\left(-3+\frac{4}{3}\varepsilon\right)}
|1-w|^{2\left(1-\frac{1}{3}\varepsilon\right)} \int d^2y |y-wx|^{2(-1+
\varepsilon)}
\en
The integration gives a result of order $O(1)$ and for the structure constant
one finds explicitely $D=\sqrt{3} + O(\varepsilon)$. With this result for the
two-point function (\ref{expan}) we shall have the following expansion for the
$\beta$-function
\eq
\beta(g)=\varepsilon g + \varepsilon A g^3 + O(g^4)
\label{beta1}
\en
where $A$ is a numerical coefficient of order $O(1)$. Again as before a naive
computation leads to a "fixed point" $g^*\sim O(1)$. We have to conclude
that if the models considered above actually describe a RG flow from $\Mt_p$ to
$\Mt_{p-1}$ as indicated by the TBA analysis, it should have an essentially
nonperturbative nature.

One remark is in order. The difference in the
central charges of the models $\Mt_p$ and $\Mt_{p-1}$ connected by the RG
trajectory is
\eq
\D c= \frac{6(3p^2+1)}{p(p^2-1)} \sim \frac{18}{p} \sim O(\varepsilon)~~\mbox
{when}~~ p\to\infty
\en
At the same time, as we argued above, there is no nontrivial fixed point
perturbatively near the origin $g=0$. One possible explanation of this
fact is as follows. The $c$-theorem of Zamolodchikov is no more valid in our
case of nonunitary models. But if we fix the "metric" $G$ in the space of the
coupling constants to a (positive or negative)
number by a suitable choice of basis, then
\eq
\D c = -12G\int_0^{g^*} \beta(g)dg
\en
Inserting the explicit expression for $\beta(g)$ (\ref{beta}) or
(\ref{beta1}) one can convince himself that it is exactly the value
$g^*\sim O(1)$ that ensures $\D c \sim O(\varepsilon)$!

\section{Conclusions speculations and generalizations}

In this paper we have learned that, allowing the perturbing parameter to take
complex values, one can greatly enrich the phenomenology of the RG space of
actions of two-dimensional Quantum Field Theory. This goes in the same
direction of some recent work of Fendley, Saleur and Al.
Zamolodchikov~\cite{FSZ1,FSZ2} on the Sine-Gordon model with imaginary coupling
(mass). In the case examined in the present paper we are lucky enough to be
able to find separately reasonable TBA systems for the description of the
massive and massless regimes of the models. This allows the very detailed
analysis done throughout the paper. In other cases, including the Sine-Gordon
one of~\cite{FSZ2}, this could not be so easy to imagine
at first glance, and one
has to resort to numerical analytic continuation (Pad\'e approximants)
to explore the possibilities in the complex $\lambda$ plane. This gives in
general less accurate results, however in most situations the precision for the
IR central charge is still good enough to establish some interesting results,
as in parafermionic theories perturbed by their generating
parafermion~\cite{progress}.

The possibility to have a massless
flow in a certain direction in the complex $\lambda$ plane
with a $\bf{Z}_2$ odd perturbing field
is restricted to the cases  with conformal dimension
$\Delta =\bar{\Delta} \geq {1 \over 2}$. This guarantees that the condition
found in~\cite{KM2} to
have square root singularities in the negative $r$ axes does not hold,
thus making it
possible to have a singularity free flow. Notice that the
condition $\Delta =\bar{\Delta} > {1 \over 2}$ holds
neither for the $\phi_{12}$ nor for the $\phi_{21}$
perturbations in the minimal unitary series.
We have therefore to disappoint our readers about exciting
possibilities like critical Ising model in imaginary magnetic field, etc...
Possible candidates for massless flows generated by $\phi_{21}$ are the
$M_{p,q}$ models with the condition
\eq
6p > 3q > 4p
\en
It should be interesting to explore which of them really admit a massless
flow for $\lambda\to i\lambda$.

Other interesting candidates for this phenomenon can be found outside the range
of minimal models. Substituting the $T_1\times A_k$ structure of TBA with a
$T_n \times A_k$, one could speculate about similar series of massive-massless
flows appearing for each $W_{n+1}$ organized non-unitary minimal models.
All these series should be related to $A_{2n}^{(2)}$ affine Toda QFT, much the
like the $\Mt_p +\psi$ are related to Izergin-Korepin. Also, many unitary
theories perturbed by a relevant
$Z_{2}$ odd field with $\D > {1 \over 2}$ could be considered for this
phenomenon. Among these, the most interesting are, perhaps, the
${G_{k} \times G_{k} \over G_{2k}}$ models, perturbed by their
$\phi^{id,id}_{adj}$ operator. The list of possible generalizations
is very long.

An important direction of investigation that should be developed is the
observation that, in analogy with what has been done in~\cite{FSZ2}, {\em
undoing the truncation} in $\Mt_p+\psi$ could lead to c-theorem violating
flows between different points of some $c=-2$ critical
line, interpreted as a kind of
``Izergin-Korepin at imaginary coupling",
exactly as reported in~\cite{FSZ2} for
the Sine-Gordon model. The possible
relevance of these flows for random walks and
polymer physics should be considered with some attention.

Another interesting issue is the violation of the Zamolodchikov $c$-theorem in
many of these flows. The function $c(t)$ ($t$ a RG ``time") that
decreases monotonically in the $c$-theorem, is not defined in the same manner
as the scaling function of TBA. However one can think that the two differ by,
say, a ``different choice of Renormalization scheme",
although this sentence has poor meaning in this context. So we expect that the
{\em qualitative} behaviour of the two are the same,
thus the conjecture that also
the TBA scaling function should decrease monotonically, which is confirmed by
{\em all} the cases of unitary theories analyzed by TBA so far.
It is a known fact that the $c$-theorem as it is, holds only
for {\em unitary} theories, no surprise then if half of the models of our
$\Mt_p + \psi$ series violate it. However people have speculated for a long
time about some $\ct$-theorem, that could hold in non-unitary theories if we
substitute the scaling function $c(r)$ with the effective one $\ct(r)$. We
observe in all our $\Mt_p+\psi$ models that such a $\ct$-theorem holds. This is
true also for all the generalizations (like $T_n \times A_k$) where one can
explicitly write a massless TBA in terms of right and left movers. The cases
that violate $\ct$-theorem too seem to us to be those justified only by
analytic continuation, but not admitting a right-left mover symmetric TBA,
which should also give a signal that $S_{+-}\not= S_{-+}$.
What is the deep physical meaning of this fact (parity violation?) should be
matter of interesting further investigation.

\vspace{.5cm}

{\bf Acknowledgements} -- We are grateful to
Ferdi\-nando Gliozzi, Ennio Quat\-trini and Galen
Sotkov for useful discussions.
Anni Koubek was very kind
in clarifying some points on S-matrix analiticity and giving
helpful hints for the numerical work.
In particular we would like to express our gratitude to Patrick Dorey for his
careful reading of the manuscript and for many interesting remarks.
R.T. thanks the Theory Groups at Bologna and Durham Universities for the
kind hospitality during various stages of this work.

\end{document}